\documentclass[trackchanges, twocolumn]{aastex7}
\usepackage{graphicx}
\usepackage{hyperref}
\usepackage{amsmath, amssymb, amsthm}

\begin{document}

\title{Magnetic Field Configuration of a Quiescent Prominence Revealed by Large-amplitude Longitudinal Oscillations in End-view Observations}

\author[0000-0003-4787-5026]{Jun Dai}
\affiliation{Key Laboratory of Dark Matter and Space Astronomy, Purple Mountain Observatory, CAS, Nanjing, 210023, China}
\affiliation{Astronomical Observatory, Kyoto University, Sakyo, Kyoto, Japan}
\email{daijun@pmo.ac.cn}

\author[0000-0002-5279-686X]{Ayumi Asai}
\affiliation{Astronomical Observatory, Kyoto University, Sakyo, Kyoto, Japan}
\email{asai@kwasan.kyoto-u.ac.jp}

\author[0000-0003-0057-6766]{Dechao Song}
\affiliation{Key Laboratory of Dark Matter and Space Astronomy, Purple Mountain Observatory, CAS, Nanjing, 210023, China}
\email{dcsong@pmo.ac.cn}

\author[0000-0002-1190-0173]{Ye Qiu}
\affiliation{Institute of Science and Technology for Deep Space Exploration, Suzhou Campus, Nanjing University, Suzhou 215163, China}
\email{qiuye@smail.nju.edu.cn}

\author[0000-0002-9121-9686]{Zhe,Xu}
\affiliation{Yunnan Observatories, Chinese Academy of Sciences, 396 Yangfangwang, Guandu District, Kunming 650216, China}
\affiliation{Yunnan Key Laboratory of Solar Physics and Space Science, 396 Yangfangwang, Guandu District, Kunming 650216, China}
\email{xuzhe6249@ynao.ac.cn}

\begin{abstract}

Prominence seismology, applied to the large-amplitude longitudinal oscillation, is used to indirectly diagnose the geometry and strength of the magnetic fields inside the prominence.
In this paper, combining imaging and spectroscopic data, the magnetic field configuration of a quiescent prominence is revealed by large-amplitude longitudinal oscillations observed in end view on 2023 December 4.
Particularly, the prominence oscillation involved blueshift velocities in Dopplergrams and horizontal motions in extreme-ultraviolet (EUV) images.
Originally, the prominence oscillation was triggered by the collision and heating of an adjoining hot structure associated with two coronal jets.
The oscillation involved two groups of signals with similar oscillatory parameters, a three-dimensional (3D) initial amplitude of $\sim$40 Mm and a 3D velocity amplitude of $\sim$48 km s$^{-1}$,
both lasting for $\sim$4 cycles with a period of $\sim$77 minutes, with a phase difference of $\sim\pi/4$.
While the angle between 3D velocities and the prominence axis ranges from 10$^\circ\mbox{--} 30^\circ$.
Two methods, utilizing time-distance diagrams and velocity fields, are employed to calculate the curvature radius of magnetic dips supporting the prominence materials.
Both methods yield similar value ranges and trends from the bottom to the top of magnetic dips,
with the curvature radius increasing from $\sim$90 Mm to $\sim$220 Mm, then decreasing to $\sim$10 Mm, with transverse magnetic field strength $\ge$25 Gauss.
From this, the realistic 3D geometry of the prominence magnetic dips is determined to be sinusoidal.
To the best of our knowledge, we present the first accurate calculation of the 3D curvature radius and geometry of the prominence magnetic dips based on longitudinal oscillatory motions.

\end{abstract}


\correspondingauthor{Jun Dai}
\email{daijun@pmo.ac.cn}

%%%%%%%%%%%%%%%%%%%%%%%%%%%%%%%%%%%%%%%%%%%%%%%%%%%%%introduction
\section{Introduction} \label{sec:intro}

Prominences or filaments are the structural features filled with cool-dense plasma materials suspended in the solar corona \citep{labrosse2010,mackay2010}.
The oscillatory motions of prominence are separated into two categories based on the velocity amplitude,
including small-amplitude oscillations ($\leq$10 km s$^{-1}$) involved a small volume of prominence materials
and large-amplitude oscillations (LAOs; $\ge$10-20 km s$^{-1}$) affecting the full prominence or most part of it \citep[see the review of][]{Arregui2018}.
Furthermore, according to the direction, the LAOs of prominence can be divided into large-amplitude transverse oscillations \citep[LATOs; e.g.,][]{isobe2006,isobe2007} and large-amplitude longitudinal oscillations \citep[LALOs; e.g.,][]{jing2003,jing2006,vrnak2007}.

LATOs of prominences are usually considered as a long uniform cylinder oscillating perpendicular to its axis, where the main restoring force is supplied by the magnetic tension force of the prominence \citep{Kleczek1969}.
LATOs are typically triggered by the interaction between Moreton waves \citep{eto2002,asai2012,liu2013} or EUV waves \citep{dai2012,zhang2018,dai2023} sweeping over and squeezing, which transfers kinetic energy.
In detail, relative to the plane of the local photosphere, LATOs can be divided into vertical and horizontal oscillations \citep[see the review of][and references therein]{Tripathi2009, Arregui2018}.
Based on previous reports, the observational characteristics of the vertical and horizontal LATOs from different viewpoints can be generally summarized as follows:
\begin{enumerate}
     \item Top-view: when observed on the disk, the line-of-sight (LOS) and plane-of-sky (POS) motions of the whole filament correspond to the nearly vertical \citep[e.g.,][]{okamoto2004, shen2014a} and horizontal \citep[e.g.,][]{shen2017} oscillations.
     Among these, the vertical LATOs are accompanied with the "winking filament" characterized by the alternative appearance and disappearance in H$\alpha$ wing images, resulting from the LOS velocity leading to the Doppler shift from H$\alpha$ center to the line wings.
     Moreover, when far away from the disk center, 3D projection effects of LATOs and the local photosphere plane are both required to be considered \citep{dai2023}.
    \item Side-view: when observed on or near the limb, LOS and POS motions correspond to the horizontal \citep[e.g.,][]{isobe2006} and vertical \citep[e.g.,][]{kim2014, zhang2016} oscillations.
    \item End-view: when observed on the limb, POS motions parallel to the local photospheric plane correspond to the horizontal oscillations \citep[e.g.,][]{liu2012, Gosain2012}. While the vertical oscillations have not been detected or reported (to our best knowledge),
    this is probably due to the leg and shoulder of the prominence blocking the oscillating signals.
\end{enumerate}

Generally, the prominence materials of LALOs are considered to oscillate along the magnetic dips in the supporting flux rope,
where the projected gravity along the dip is dominant for the restoring force.
Prominence seismology based on this can effectively estimate the curvature radius and transverse magnetic field strength of the dips \citep{luna2012,zhang2012AA,lunak2012}.
LALOs are frequently triggered by solar flares or subflares \citep{jing2003,jing2006,vrnak2007,lizhang2012,zhang2012AA,zhang2017a,zhang2020}, as well as coronal jets \citep{luna2014,zhang2017b,luna2021,ni2022} adjacent to the footpoints of the prominences where the plasma is subjected to evaporation,
resulting in hot flow to push the cold-dense materials away from the initial position \citep{lunak2012,luna2012}, and sometimes are induced by the impact of coronal shock waves in a direction roughly along the prominence axis \citep{shen2014b,pant2016,pan2025} or by the merging of two adjacent filaments \citep{luna2017}.

Furthermore, in most observational cases \citep{luna2018} and 3D scenario \citep{zhou2018}, there is a certain angle ($\leq 40^\circ$) between the magnetic field of the dips supporting LALOs and the axis of flux ropes.
Therefore, the LALOs observed from different viewpoints also appear different distinctions:

 \begin{enumerate}
     \item Top-view: which are most frequently observed on the disk, the materials roughly oscillate along the spine of filaments, but have a certain angle ranging from $10^\circ$ to $30^\circ$ \citep[e.g.,][]{luna2014,zhang2017b,luna2018}.
     \item Side-view: which are rarely observed on the limb, the materials also roughly move along the horizontal axis of the prominence\citep[e.g.,][]{zhang2017a,ouyang2020} with a slight Doppler shift \citep{zapior2019}. While in the high-resolution images,
     the threads and their oscillating trajectories were aligned with the concave-outward magnetic dips\citep{zhang2012AA}.
     \item End-view: which has not been reported currently, but simultaneous POS and LOS motions can be predicted, and the Doppler velocity is significantly greater than the POS velocity.
 \end{enumerate}

Both LALOS and LATOS involve the lateral displacement, which means that we can not simply rely on the lateral displacement to confirm that prominence oscillations are of transverse mode.
Meanwhile, the components of LALOs in two directions theoretically have the same oscillation parameters \citep{chen2017}, while the period, phase, or damping time of simultaneous LALOs and LATOs in a single filament or prominence have obvious differences, which could be distinguished from the top view or side view \citep{zhang2017b,dai2021,tan2023}.
However, LATOs and LALOs observed from the end view both involve evident POS motions, thus, the combination of Doppler velocity is required to obtain a 3D oscillatory direction to determine whether LAOs are longitudinal or transverse, or simultaneous.

In addition, although the two-dimensional (2D) simplified pendulum model has been widely applied on LALOs, the realistic geometry of the magnetic dips is not semicircular \citep{luna2022}.
Theoretically, the geometry and curvature radius of the dips can be directly obtained from the evolution of 3D velocity, including the values and direction.
However, such investigation remains lacking due to the limitations in the temporal and spatial resolution of the data.

On 2023 December 04, a large-amplitude longitudinal oscillation of a prominence on the northwestern limb was detected in EUV images and H$\alpha$ image spectra, which provides us a rare opportunity to study the LALOs from the end view.
In this paper, we focus on the 3D direction and velocities of the LALOS, as well as the accurate calculation of the 3D curvature radius and geometry of the prominence magnetic dips.
The observation and data analysis are described in Section~\ref{sec:obs}.
The results are presented in Section~\ref{sec:res}.
A comparison with previous findings and a brief conclusion are given in Section~\ref{sec:sum}.

%%%%%%%%%%%%%%%%%%%%%%%%%%%%%%%%%%%%%%%%%%%%%%%%%%%%%%observation
\section{Observation and data analysis} \label{sec:obs}

\subsection{Observation and data}
On 2023 December 04, the large-amplitude oscillation of a quiescent prominence located at the northwestern limb was simultaneously recorded by H$\alpha$ Imaging Spectrograph (HIS) on board the Chinese H$\alpha$ Solar Explorer \citep[CHASE;][]{Lichuan2022}
and Atmospheric Imaging Assembly \citep[AIA;][]{lemen12} on board the Solar Dynamics Observatory \citep[SDO;][]{Pesnell2012}.
Since CHASE was launched into a Sun-synchronous orbit, it is necessary to note that the periods of CHASE observation for this event are 12:25-12:45 UT, 14:00-14:25 UT, and 15:35-15:55 UT, respectively.

We employed the full-disk images in seven EUV (94, 131, 171, 193, 211, 304, and 335{~\AA}) wavelengths obtained from AIA to investigate the dynamics of the prominence in the plane of sky.
The level\_1 data from AIA with a cadence of 12 s and a spatial resolution of 1$\farcs$2 were calibrated using the standard Solar SoftWare (SSW) programs \texttt{aia\_prep.pro}.
Moreover, we also employed the full-disk spectroscopic observations with a temporal resolution of about 70 s at H$\alpha$ (6559.7-6565.9{~\AA}) wavebands obtained from HIS in raster scanning mode (RSM).
The Level-1.5 data products involved dark field correction, flat-field and slit-image curvature correction, wavelength, and
intensity calibration, and coordinate transformation
\citep{Qiuye2022SCPMA}.
In normal observation, the full-disk H$\alpha$ images have a spatial resolution of 0$\farcs$52 pixel$^{-1}$ and a spectral resolution of 0.024{~\AA} pixel$^{-1}$.
While the data employed in this study were obtained in binning mode, where the full-disk H$\alpha$ images have a spatial resolution of 1$\farcs$04 pixel$^{-1}$ and a spectral resolution of 0.048{~\AA} pixel$^{-1}$.
In addition, a more updated introduction for CHASE data can be viewed on the following website: \text{https://ssdc.nju.edu.cn.}

\subsection{Models and Methods}
The cloud model \citep{Beckers1964,Mein1988,chae2014,hong2014} is widely used to indirectly derive the 3D velocity of erupting or oscillating filaments based on H$\alpha$ observations \citep{Mor2003,Mor2010,shen2014a,dai2023,qiuye2024}.
To obtain the Doppler velocity, we first define the line center of the averaged spectral profile in a quiescent region [700$\arcsec$, 715$\arcsec$]$\times$[575$\arcsec$, 585$\arcsec$] as zero shift to avoid the influence of solar rotation and instrument movement.
Different from filaments, the prominence can be regarded as a luminous slab against a dark background. To acquire its line-of-sight velocity, we recognize the prominence region by emission line profiles,
and then adopt the single cloud model without the absorption term of bright background to fit the spectral line profile \citep{Li2019, Yu2020}.

The POS velocity fields of the oscillating prominence in EUV images were calculated using the dense optical flow method in the \textit{OpenCV} package (\textit{cv2.calcOpticalFlowFarneback}), which is based on Gunnar Farnebäck's algorithm \citep{Gunnar2003}.
Optical flow describes the motion of brightness or intensity patterns across consecutive images.
The tracking of motions is predicated on assumptions of brightness constancy, spatial coherence, and small motion.
This method has been applied to high-resolution TiO images to derive the motions ($v_x$ and $v_y$) on the solar surface \citep{xuzhe2025}.
In this study, the optical flow method was empirically applied to AIA EUV images.
In detail, calculation was carried out with a 5 pixel window (computing displacement of each pixel by tracking 5$\times$5 pixels around it at the next moment), corresponding to a spatial scale of $\sim$3$\farcs$0, which was determined to maintain spatial detail.
The minimum detectable velocity here is $\sim$2 km s$^{-1}$, since the method could track sub-pixel motions.
The velocity field of each time period involves 6 consecutive EUV images to improve the signal-to-noise ratio.

%%%%%%%%%%%%%%%%%%%%%%%%%%%%%%%%%%%%%%%%%%%%%%%%%%%%%%%%%%result

\begin{figure*}
\centering\includegraphics[width=0.95\textwidth]{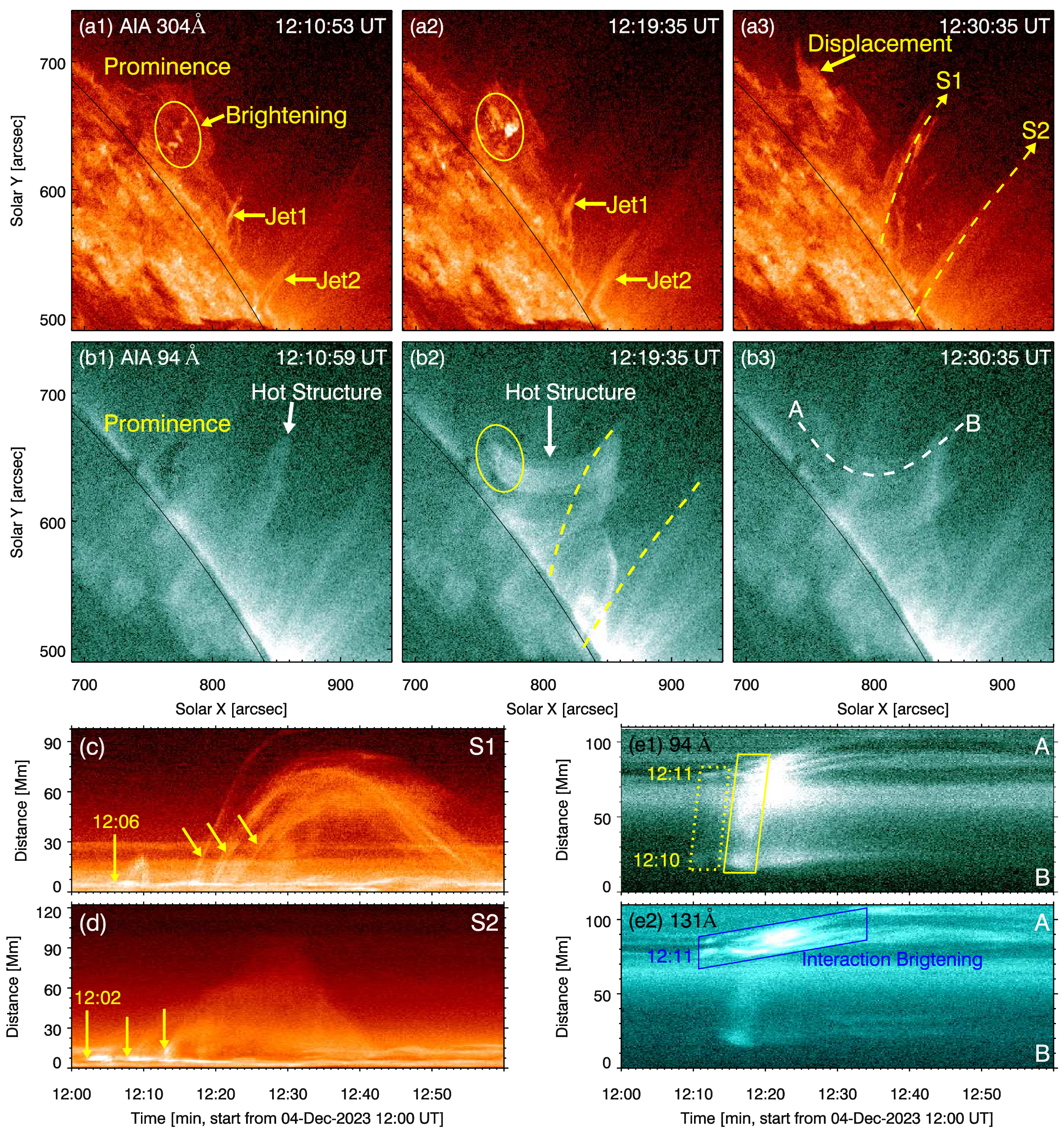}
%\plotone{figure1.eps}
\caption{\small{The triggering process of the prominence oscillation.
Panels (a1-a3) give three EUV images in AIA 304{~\AA}.
The yellow arrows point to the coronal jets and the displacement of the prominence materials, respectively.
Panels (b1-b3) give three EUV images in AIA 94{~\AA}.
The white arrows point to the hot structure.
The yellow ellipses in panel (a1-a2) and (b2) mark the brightening region inside the prominence.
The yellow dashed lines in panels (a3-b2) represent the rising track of the two coronal jets.
Panels (c-d): Time-distance diagrams of the slices S1 and S2 in AIA 304{~\AA}.
The yellow arrows in panels (c-d) mark the start times and different phases of the coronal jets;
Panels (e1-e2): Time-distance diagrams of the slice AB in AIA 94{~\AA} and 131{\AA}.
The two yellow parallelograms in panel (e1) mark the propagation of the hot structure.
The blue parallelogram in panel (e2) outlines the interaction brightening between the hot structure and the prominence.
(An animation of this figure is available.)}}
\label{fig1}
\end{figure*}

\section{Result} \label{sec:res}

\subsection{Triggering of Prominence Oscillation}

In Figure~\ref{fig1}, the top two rows of panels show EUV images in 304{~\AA} and 94{~\AA}, respectively.
A quiescent prominence was clearly seen in multi-wavelength images.
Accompanied by the appearance of two adjoining coronal jets \citep{Shibata1994} and a hot structure only detected in 94{~\AA} and 131{~\AA} in the south-west of the prominence, the inside brightening marked by yellow ellipses in Figure~\ref{fig1}(a1-a2, b2) appeared in the prominence,
followed by a north-east displacement of prominence materials marked by a yellow arrow in Figure~\ref{fig1}(a3).

In order to investigate the kinematics of the two coronal jets, the curved slices S1 (with a distance of 98 Mm, corresponding to the rising track of Jet 1) and S2 (with a distance of 122 Mm,
corresponding to the rising track of Jet 2) were selected to construct the time-distance diagrams in 304{~\AA} in Figure~\ref{fig1}(c-d).
Meanwhile, to investigate the interaction between the hot structure and the prominence, the curved slice AB with a distance of 110 Mm was selected to construct the time-distance diagrams in 94{~\AA} and 131{~\AA} in Figure~\ref{fig1}(e1-e2).

It can be seen that Jet 1 and Jet 2 started to rise at $\sim$12:06 UT and $\sim$12:02 UT, respectively.
The two coronal jets were both recurrent, where Jet 1 and Jet 2 exhibited four and three segments, which were marked by the yellow arrows in Figure~\ref{fig1}(c-d).
The plasma flows in the two coronal jets exhibited a parabolic trajectory, consistent with the dynamical characteristics in the previous reports \citep{liu2009,lu2019,huang2020,zhang2021R}.

Furthermore, as shown in Figure~\ref{fig1}(e1-e2), the initial trace of the hot structure occurred at $\sim$12:10 UT in the vicinity of the B end.
Approximately one minute later, at around 12:11 UT, the brightening appeared inside the prominence, corresponding to the remote brightening of the hot structure.
As delineated by the blue parallelogram, the interaction brightening between the hot structure and prominence material persisted for $\sim$23 minutes.
As marked by the two yellow parallelograms, the remote brightening appeared from $\sim$12:10 UT to $\sim$12:14 UT, while the propagating manifestation of hot materials inside the hot structure was detected from $\sim$12:15 UT to $\sim$12:20 UT.
Considering that no solar flares were recorded in this event,
the energy propagation of a heat conduction\citep{shimojo2007} and heat flux\citep{wangliu2012} along the magnetic field of the hot structure could dominate the remote brightening caused by heating the prominence material.

Combining Figure~\ref{fig1}(e1-e2) and Figure~\ref{fig2}(c),
the energy propagation from the hot structure to the prominence materials began at $\sim$12:11 UT and immediately induced a little brightening plasma to deviate.
While the cold and dense prominence materials began to deviate from their initial position at $\sim$12:15 UT,
concurrent with the observed propagation of hot materials inside the hot structure,
causing the increase of gas pressure inside the flux rope at the side of the brightening.
These two phases indicated that the displacement of the prominence materials was triggered by the combined effects of the collision and continuous heating.

As shown in Figure~\ref{fig1}(b2), the rooted locations of the two coronal jets and the hot structure were highly correlated in the plane of sky,
even considering the projection effect.
However, the physical correlation between the hot structure and the two coronal jets can not be obtained due to the lack of multi-view observations and the photospheric LOS magnetograms.

Nevertheless, spatially and temporally, the formation of the hot structure could be closely associated with the two coronal jets.
Therefore, it is suggested that the prominence oscillation was triggered by the collision and heating of the adjoining hot structure associated with two coronal jets.

\subsection{Large-amplitude Longitudinal Oscillation} \label{LALO}

\begin{figure*}
\centering\includegraphics[width=0.85\textwidth]{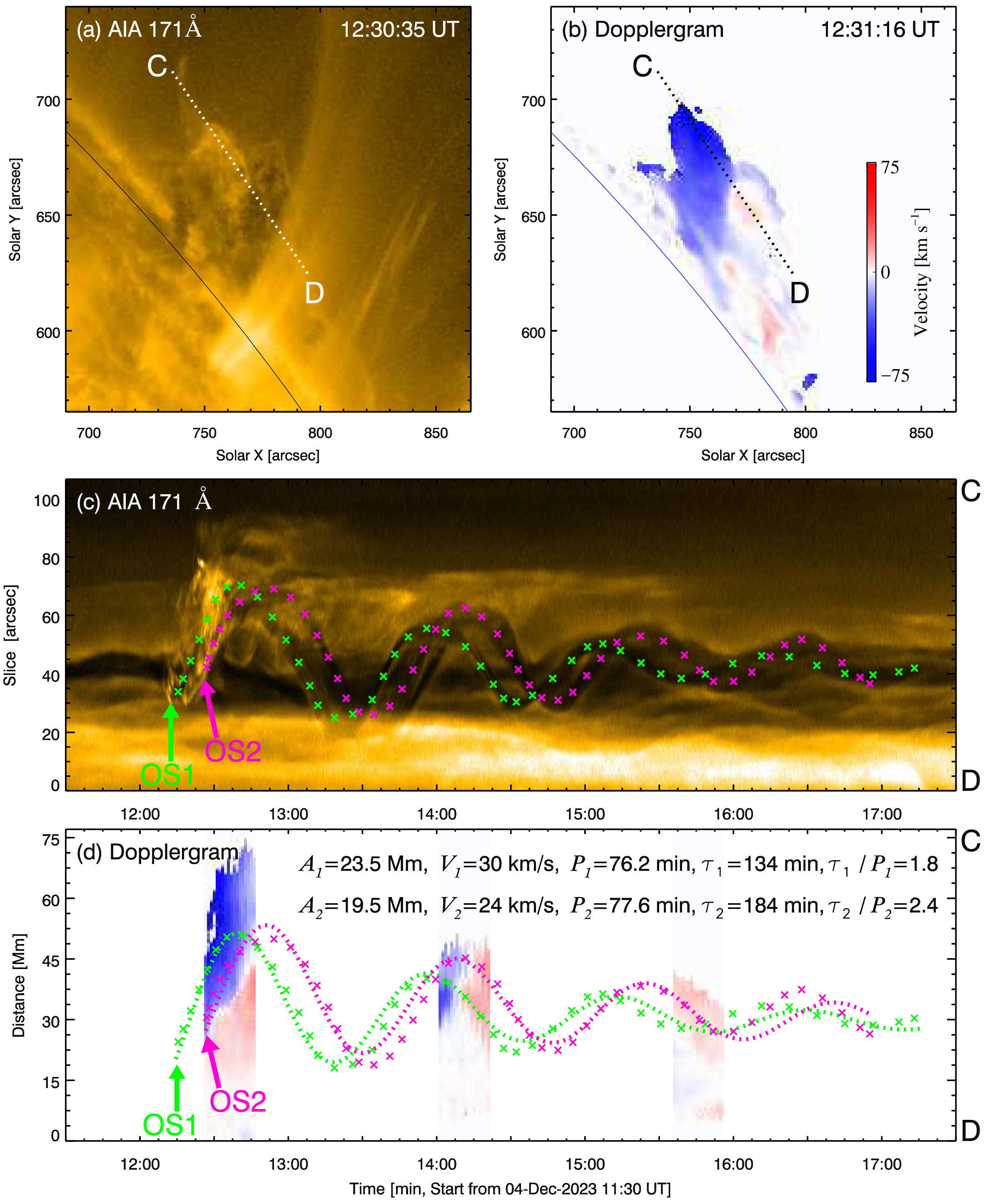}
%\plotone{figureos.eps}
\caption{\small{Panel (a): An EUV image in AIA 171{~\AA} of the oscillating prominence.
Panel (b): A Dopplergram showing the velocities of the oscillating prominence materials.
Panel (c): Time-distance diagrams of the slice CD in AIA 171{~\AA}.
The green and magenta crosses outline the prominence oscillations OS1 and OS2 between 11:30 UT and 17:30 UT, respectively.
Panel (d): Time-distance diagrams of the slice CD in Dopplergrams,
where the three periods are 12:25 - 12:45 UT, 14:00 - 14:25 UT, 15:35 - 15:55 UT, respectively.
The extracted positions of OS1 and OS2 in panel (c) are overlaid with green and magenta crosses,
and the fitted curves are overlaid with green and magenta dotted lines, respectively.
Corresponding parameters are labeled.
(An animation of this figure is available.)}}
\label{fig2}
\end{figure*}

After interacting with the hot structure, from $\sim$12:15 UT,
the obvious horizontal oscillations and Doppler velocities of the prominence materials were respectively detected in EUV images and Dopplergrams, as shown in Figure~\ref{fig2}(a-b), indicating that the oscillation contained both POS and LOS velocity components ($V_{POS}$ and $V_{LOS}$) at the same time.

To investigate the kinematics of the prominence involved two velocity components, we select a straight slice CD with a length of 77 Mm to construct the time-distance diagrams with 171{~\AA} and Dopplergrams, which were plotted in Figure~\ref{fig2}(c-d).
Expressly, two groups of oscillating signals (OS1 and OS2) were visible in Figure~\ref{fig2}(c), which are damping and lasting for about 4 cycles, and OS2 lags behind OS1 for $\sim$10 minutes.
While the time-distance diagrams of Dopplergrams covered approximately one quarter of each cycle during the first three cycles of OS1 and OS2.

To accurately determine the parameters of the prominence oscillation, we fit the green and magenta curves in Figure~\ref{fig2}(c) using \texttt{mpfit.pro} and the following exponentially decaying sine function:
\begin{equation} \label{eqn-1}
  A(t) = A_{0}\sin (\frac{2\pi t}{P} + \psi )e^{-\frac{t}{\tau}} + A_{1}t + A_{2},
\end{equation}
where $P$ is the period, $\tau$ is the damping time, $\psi$ is the initial phase,
while the $A_0$ is the initial POS amplitude, $A_{1}t+A_{2}$ represents a linear term of the equilibrium position of the prominence.

In Figure~\ref{fig2}(d), the green and magenta crosses represent the extracted positions of OS1 and OS2,
 the fitting result using Equation~\ref{eqn-1} was overlaid with green and magenta dashed lines.
As seen from the fitted parameters listed in Figure~\ref{fig2}(d), OS1 and OS2 have similar oscillatory parameters but a phase difference of $\sim \pi/4$,
an initial POS amplitude of $\sim$23.5, 19.5 Mm, and a velocity amplitude of $\sim$30, 24 km s$^{-1}$.
The period and the damping time are $\sim$76.2, 77.6 minutes and $\sim$134, 184 minutes, and the corresponding quality factor ($\tau/P$) is $\sim$1.8, 2.4.
Furthermore, the evolution of $V_{POS}$ with time can be expressed as the first derivative function of Equation~\ref{eqn-1}:

\begin{equation} \label{eqn-2}
V(t) = A_0 e^{-\frac{t}{\tau}} ( \frac{2\pi}{P} \cos ( \frac{2\pi t}{P} + \psi ) - \frac{1}{\tau} \sin ( \frac{2\pi t}{P} + \psi ) )
\end{equation}

where the parameters are the same as those in Equation~(\ref{eqn-1}).
It can be seen that the phase changes of the two sets of oscillation signals and the time-distance diagrams of Dopplergrams are almost consistent.
For the periods of 12:25 - 12:45 UT and 15:35 - 15:55 UT, the moving direction of OS1 and OS2 is the same and corresponds to the coincident blue-shift and red-shift signals, respectively.
During the period of 14:00 - 14:25 UT, the moving directions of OS1 and OS2 are opposite and correspond to the alternating blue-shift and red-shift signals, respectively.
This indicates that the oscillations are purely longitudinal or transverse, rather than simultaneous LALOs and LATOs.
To confirm whether the prominence oscillation is transverse or longitudinal, the angle between the 3D velocity ($V_{3D}$) of the oscillation and the axial direction of the prominence must be accurately calculated.

\begin{figure*}
\centering\includegraphics[width=0.99\textwidth]{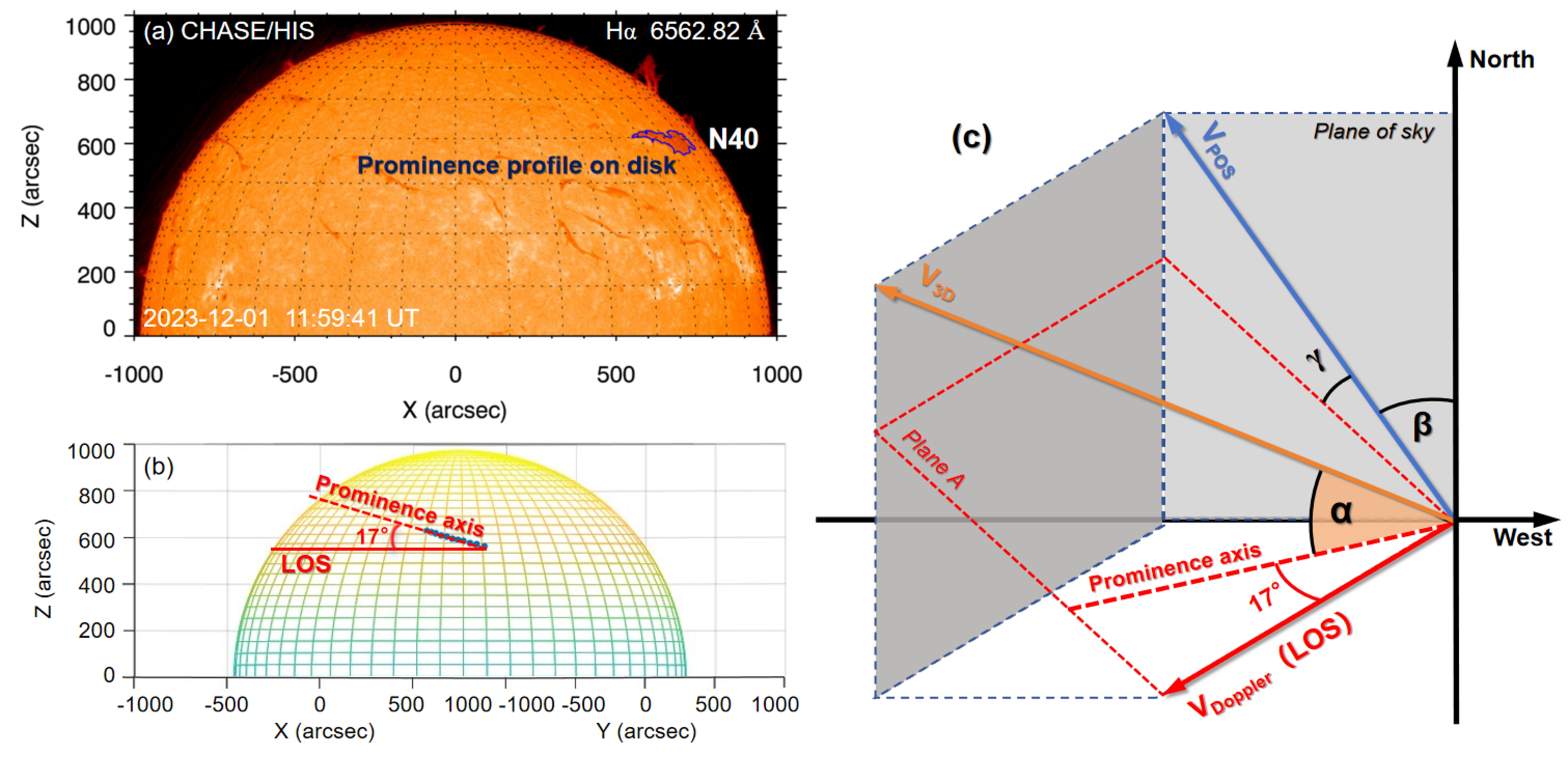}
%\plotone{figure-V.pdf}
\caption{\small{The axial direction of the prominence and the velocities of the oscillation.
Panel (a) gives an H$\alpha$ line-center image in 6562.82{~\AA} observed by CHASE on 2023 December 01 at $\sim$12:00 UT.
The blue contour outlines the profile of the prominence when it was on the solar disk.
The blue solid dots in panel (b) represent the extracted location of the prominence spine from panel (a),
and have been rotated to the center of the solar disk (corresponding to the top view) in the coordinate system.
The red dashed and solid lines mark the axial direction of the prominence spine and the LOS direction when they are projected on the disk, respectively.
Panel (c) gives a schematic diagram showing the angles between the prominence axis and oscillatory velocities.
The gray rectangle represents the plane of sky.
POS west, north, and LOS directions are taken as x-, y-, and z-axes.
The blue, red, and orange solid arrows indicate the POS velocity, Doppler velocity, and 3D velocity of oscillatory materials, respectively.
The Doppler velocity and the prominence axis are coplanar in the red rectangle (Plane A), which is perpendicular to the plane of sky.
The angle between POS velocity and y-axes, Plane A are marked as $\beta$ and $\gamma$, respectively.
$\alpha$ represents the angle between the prominence axis and the direction of the 3D velocity.}}
\label{fig3}
\end{figure*}

It can be seen from Figure~\ref{fig3}(a) that the profile of the prominence when it was on the solar disk, on December 01, was outlined by a blue contour.
The extracted location of the prominence spine and its axial direction, assuming parallel to the local photospheric plane, marked by the blue solid dots and red dashed line in Figure~\ref{fig3}(b) was projected onto the coordinate system, and was rotated to the center of the solar disk, corresponding to the top view of the prominence.
While the LOS direction was projected onto the solar disk, it was parallel to the latitude lines in the coordinate system, marked by the red solid line.
Accordingly, the angle between the prominence axis and LOS direction was determined to be $\sim$17$^\circ$.

\begin{figure*}
\centering\includegraphics[width=0.85\textwidth]{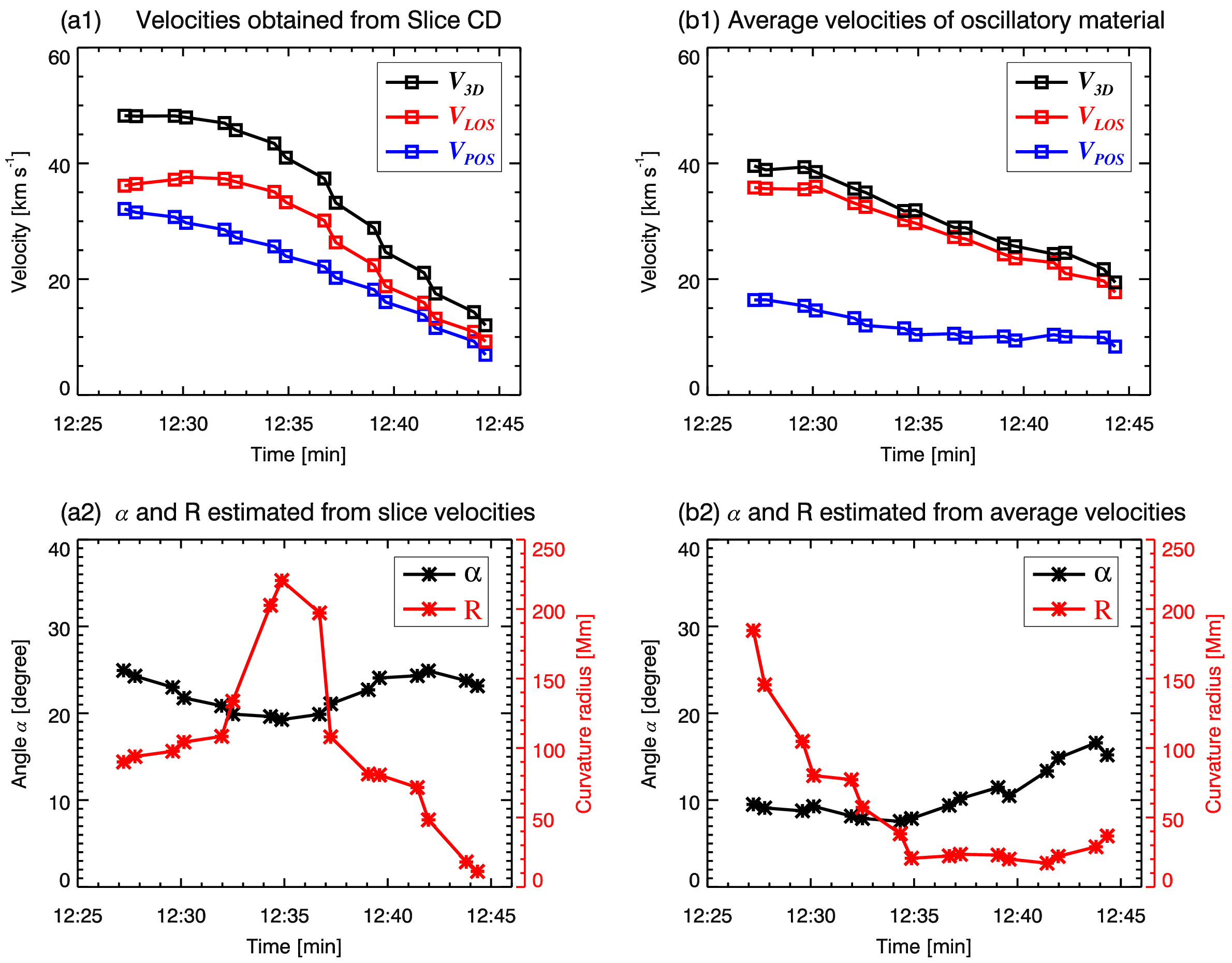}
\caption{\small{Panels (a1-b1): Velocity-time plot of $V_{POS}$, $V_{LOS}$, $V_{3D}$.
The values in panel (a1) were obtained from slice CD,
while the values in panel (b1) were obtained from the simultaneous POS velocity fields and Dopplergrams of the oscillatory prominence materials.
Panels (a2-b2): Evolution of the angle $\alpha$ and local curvature radius of magnetic dip,
where the values were estimated from the velocities in panels (a1) and (b1), respectively.}}
\label{fig4}
\end{figure*}

As illustrated in Figure~\ref{fig3}(c), a triangulation procedure of the prominence axis and oscillatory velocities is schematically represented,
in which $\alpha$ represents the angle between $V_{3D}$ and prominence axis.
The prominence axis and $V_{LOS}$ are coplanar in Plane A, which is perpendicular to Plane of sky and tangent to the solar surface.
The angle between Plane A and North direction, namely the sum of the angle $\beta$ and $\gamma$, is determined to be 40$^\circ$ based on the latitude of N40 where the prominence located.
Thus, the angle $\alpha$ can be derived from geometric relationships and trigonometric functions as follows:
\begin{equation} \label{eqn-3}
  \cos\alpha =  \frac{V_{LOS}\cos17^\circ +V_{POS} \sin17^\circ \cos \gamma} {V_{3D}}
\end{equation}
where $V_{3D}=\sqrt{V^2_{POS}+V^2_{LOS}}$,
and $\gamma=40^\circ-\beta$, represent the angle between $V_{POS}$ and Plane A.
When employing $V_{POS}$ of OS2 derived from the time-distance diagrams in Figure~\ref{fig2} and Equation~\ref{eqn-2},
the value of angle $\beta$ is determined to be 30$^\circ$ by the direction of Slice CD.
Using $V_{POS}$ of OS2 obtained from Equation~\ref{eqn-2} (the blue curve in Figure~\ref{fig4}(a1)) and the average blue-shift values neighboring to the positions of OS2 (the red curve in Figure~\ref{fig4}(a1)), the evolution of $V_{3D}$ and angle $\alpha$ within approximate first quarter period was obtained and indicated by black marks in Figure~\ref{fig4}(a1-a2).
It can be seen that the values of $\alpha$,
the angle between the oscillatory direction and prominence axis, approximately range from 20$^\circ \mbox{--}$ 30$^\circ$.

Therefore, the quiescent prominence underwent a large-amplitude longitudinal oscillation (LALO) with 3D initial amplitude of $\sim$40 Mm and 3D velocity amplitude of $\sim$48 km s$^{-1}$, while OS1 and OS2 are suggested to be oscillating signals in two adjoining magnetic dips inside the prominence.

\subsection{Curvature Radius and Geometry of Magnetic Dips}

\subsubsection{Prominence seismology based on pendulum model}

As mentioned in section~\ref{sec:intro},
the pendulum model is the effective method of indirect diagnosis and has been widely applied to the LALOs.
When applying the prominence seismology on the large-amplitude longitudinal oscillations,
two simplifying assumptions are usually proposed for the pendulum model in the previous studies \citep{lunak2012,luna2012,zhang2012AA}:
\begin{enumerate}
\item The geometry of the magnetic dips of the flux ropes supporting the prominence materials is semicircular;
\item The gravity of the oscillating materials is uniform and regardless of their height.
\end{enumerate}
In this case, the period of the oscillation can be simplified expressed as:

\begin{equation}\label{eqn-4}
  P=2\pi\sqrt{\frac{R}{g_{\odot}}},
\end{equation}

where $R$ is the curvature radius of the dip and $g_{\odot}$ =0.274 km s$^{-2}$ is the gravitational acceleration at solar surface.
Besides, the strength of the transverse magnetic field of the dips can be estimated by \citep{luna2014}:

\begin{equation}\label{eqn-5}
  B_{tr}[G] \ge (17 \pm 9)P [hr],
\end{equation}

using the observed value of $P \approx 77$ min,
the curvature radius was determined to be $\sim146$ Mm,
and the transverse strength of magnetic dips was estimated to be $22 \pm 11.5$ Gauss.

However, the gravity is not uniform when the value of the curvature radius relative to the solar radius cannot be ignored.
Thus, for the corrected pendulum model considering non-uniform gravity \citep{luna2022},
the period of the oscillation and the corresponding transverse strength of magnetic dips can be expressed as:

\begin{equation} \label{eqn-6}
 P = 2\pi \sqrt{ \frac{1}{ g_{\odot} \left( \frac{1}{R} + \frac{1}{R_{\odot}} \right) } },
\end{equation}

\begin{equation}\label{eqn-7}
B_{g}[G] \ge \frac{B_{tr}} {\sqrt{1 - \left( \frac{P}{P_{\odot}} \right)^{2}}}[hr]
\end{equation}

where $R$ and $g_{\odot}$ are same to that in Equation~(\ref{eqn-4}),
and $R_{\odot}$ = 696.3 Mm is the solar radius which is the distance from the center of the Sun to the photospheric boundary\citep{qu2013},
while $P_{\odot}=2\pi \sqrt{R_{\odot}/g_{\odot}}$=167 min is the theoretical maximum value of period, corresponding to a curvature radius of the magnetic dip approaching $R_{\odot}$.
Similarly, using the observed value of $P \approx 77$ min,
the curvature radius and transverse strength considering the non-uniform gravity are determined to be $\sim185$ Mm and $25 \pm 13$ Gauss.

\subsubsection{Estimation based on time-distance diagrams}

\begin{figure}[b]
\centering\includegraphics[width=0.45\textwidth]{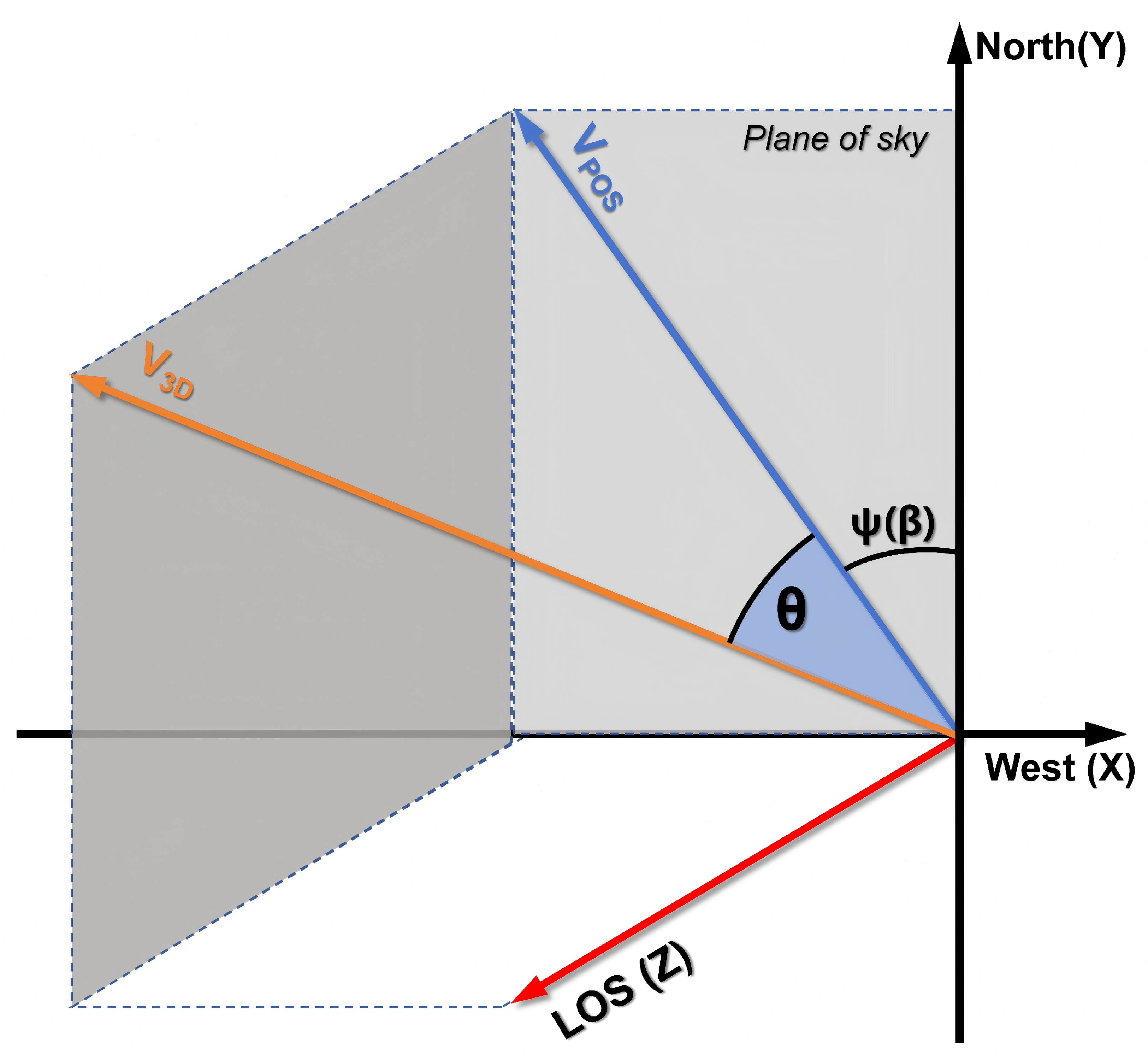}
%\plotone{figure-V.pdf}
\caption{\small{The elevation angle $\theta$ and azimuth angle $\psi$ ($\beta$ in Figure~\ref{fig3} ) of $V_{3D}$ in right-hand coordinate system.}}
\label{fig5}
\end{figure}

\begin{figure*}
\centering\includegraphics[width=0.875\textwidth]{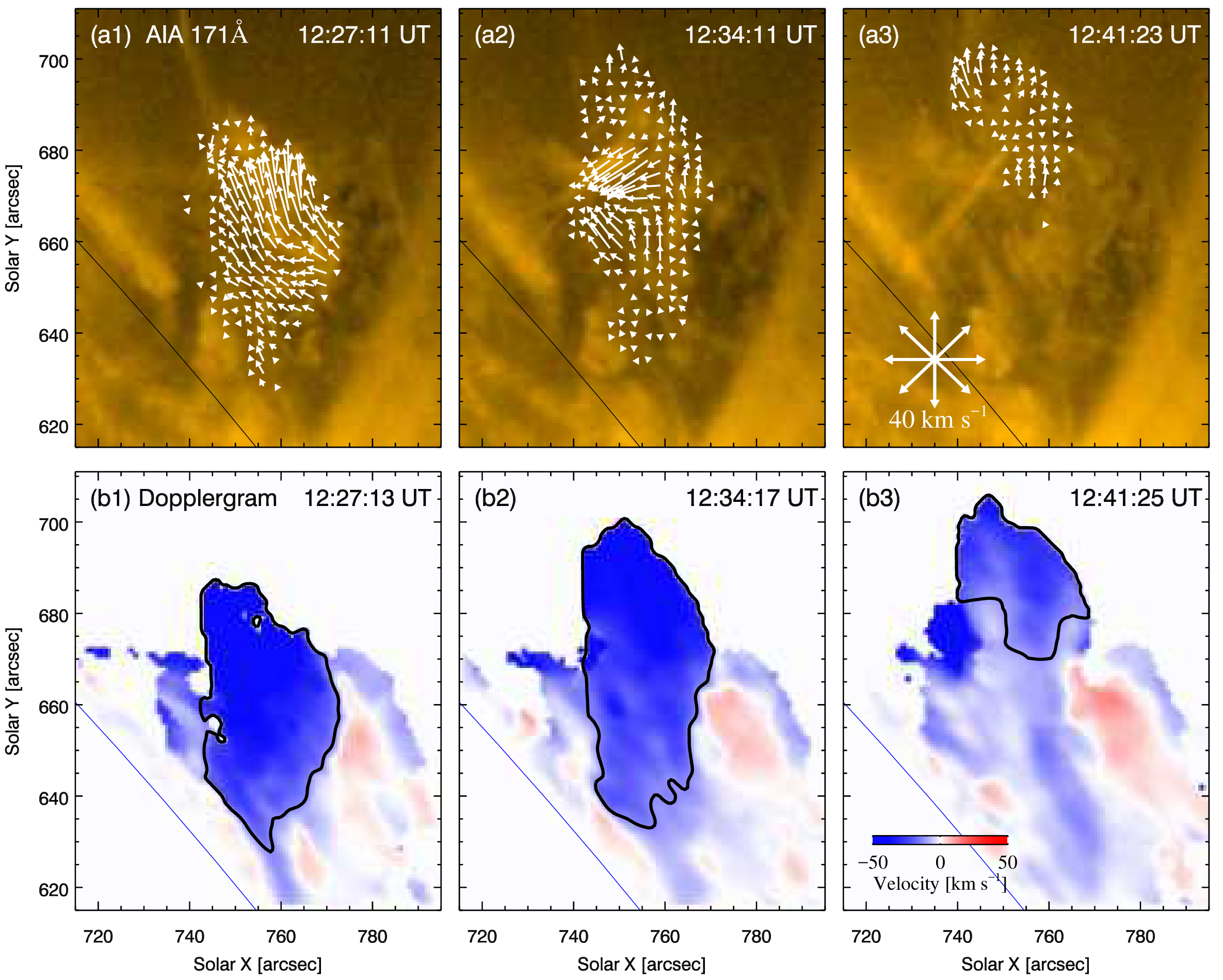}
%\plotone{figure-V.pdf}
\caption{\small{The POS velocity fields and Dopplergrams of the oscillatory prominence.
Panels (a1-a3) give three EUV images in AIA 171{~\AA} overlaid by simultaneous POS velocity fields of the oscillatory prominence materials.
Panels (b1-b3) give three Dopplergrams where the black contours outline the profile of oscillatory prominence materials. }}
\label{fig6}
\end{figure*}

Since prominence materials move along the magnetic dips, we attempted to directly calculate the 3D geometry of the dips using the evolution of the $V_{3D}$ involved values and directions.
According to the derivation in Appendix, the instantaneous 3D curvature radius can be expressed as:

\begin{equation} \label{eqn-8}
 R(t) = \frac{V_{3D}(t)}{\sqrt{
    \left( \frac{\mathrm{d}\theta}{\mathrm{d}t} \right)^{2} +
    \left( \frac{\mathrm{d}\psi}{\mathrm{d}t} \cos \theta(t) \right)^{2}
}}
\end{equation}

where $\theta$ and $\psi$ represents the elevation and azimuth angle of $V_{3D}$ in coordinate system in Figure~\ref{fig5}.
Correspondingly, $\theta(t) = arctan(\frac{V_{LOS}} {V_{POS}})$ is the instantaneous angle between $V_{3D}$ and x-y plane, and $\psi = \beta$ is the angle between the projection of $V_{3D}$ onto x-y plane and the y-axes.

Considering that the direction of $V_{POS}$ is along the Slice CD,
the angle $\beta = 30^\circ$ is unchanged.
Thus, the motion trajectory curve obtained from $V_{3D}$ can actually be regarded as two-dimensional in the plane containing of $V_{POS}$ and $V_{LOS}$.
In this case, according to the derivation in Appendix 2.2,
Equation~\ref{eqn-8} can be reformulated as:

\begin{equation} \label{eqn-9}
 R(t) = \frac{V_{3D}(t)}{\left| \frac{\mathrm{d}\theta}{\mathrm{d}t} \right|},
\end{equation}

where $\theta = arctan(\frac{V_{LOS}}{V_{POS}})$ represents the angle between $V_{3D}$ and $V_{LOS}$, $\frac{\mathrm{d}\theta}{\mathrm{d}t}$ represents the angular velocity of $V_{3D}$.
Using the variations of $V_{POS}$ and $V_{LOS}$ in Figure~\ref{fig4}(a1),
and applying the angular velocity calculation for non-uniform time intervals (Appendix 2.3),
the corresponding variations of curvature radius R are obtained and plotted in Figure~\ref{fig4}(a2) with red marks.
It can be seen that the curvature radius of the magnetic dip supporting OS2 increases from $\sim$90 Mm at its bottom to $\sim$220 Mm in its middle, then decreases to $\sim$10 Mm at its top.
This tendency indicates that the geometry of the magnetic dip is sinusoidal rather than semicircular.

\begin{figure*}
\centering
\begin{minipage}{0.495\textwidth}
  \centering
  \includegraphics[height=8cm]{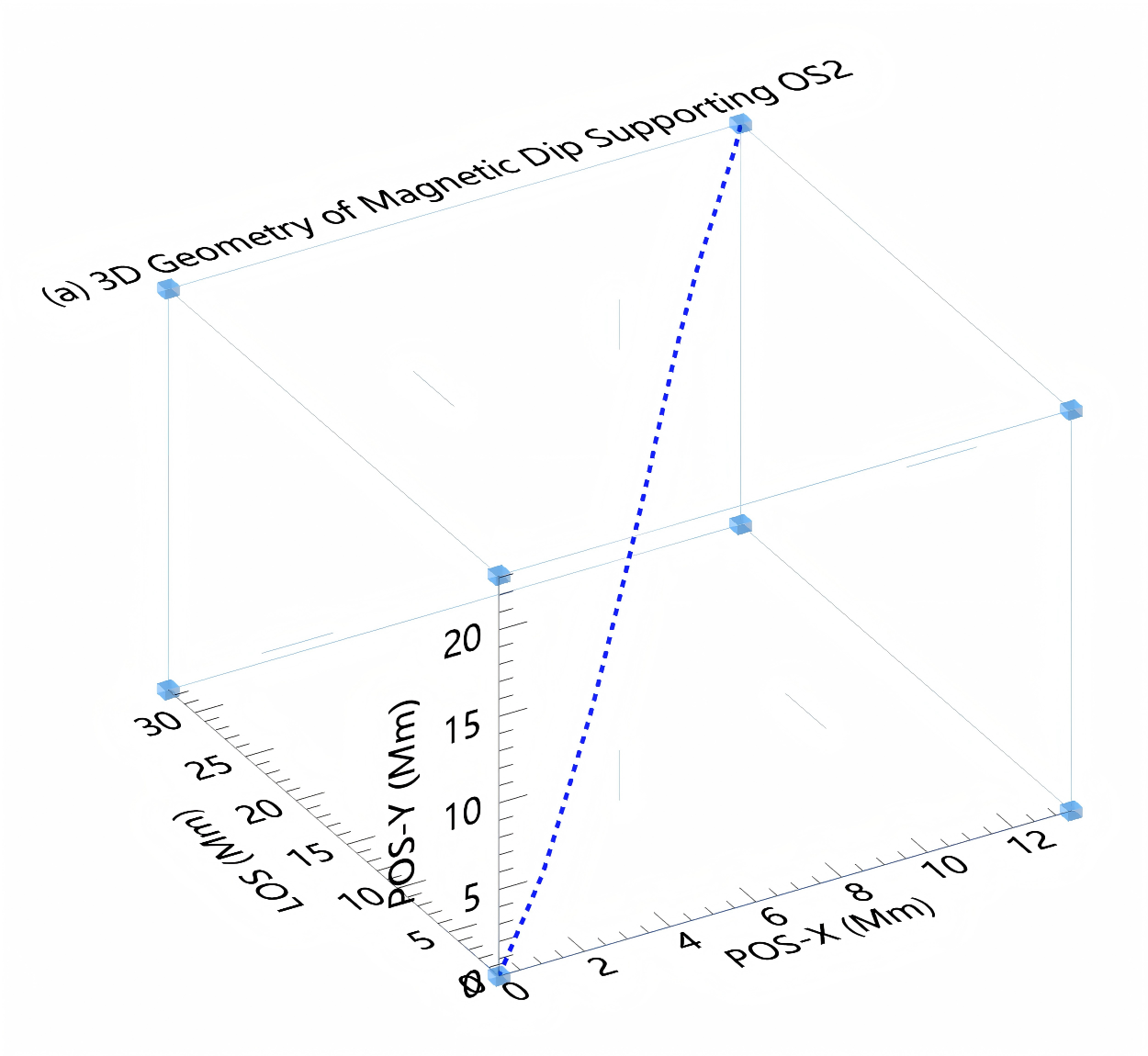}
\end{minipage}
\hfill
\begin{minipage}{0.495\textwidth}
  \centering
  \includegraphics[height=8cm]{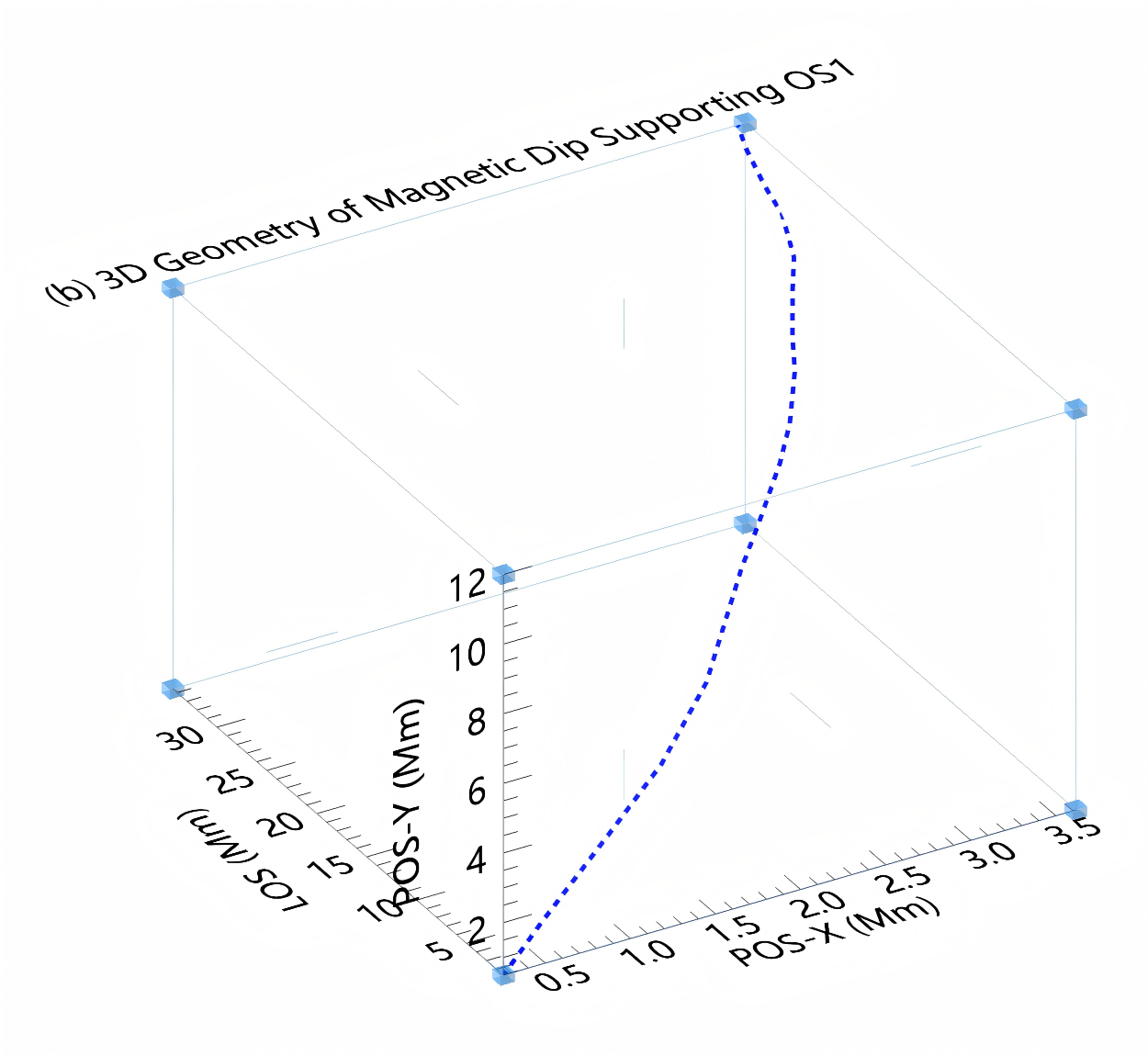}
\end{minipage}
\caption{\small{3D geometry of magnetic dips supporting OS2 and OS1 obtained from velocities.
POS north, east  and LOS directions are taken as POS-Y, POS-X and LOS axes.
Blue dotted lines represent the motion trajectories (the geometry of magnetic dips) of OS2 and OS1.}}
\label{fig7}
\end{figure*}

\subsubsection{Measurement based on velocity fields and Dopplergrams}

In order to validate the geometry of magnetic dips supporting oscillatory prominence materials inferred from $V_{3D}$ based on the time-distance diagrams, we carried out POS velocity fields by using optical flow method \citep{Gunnar2003}, as well as the Dopplergrams by using cloud model \citep{Beckers1964}, thereby obtaining $V_{3D}$ of the oscillatory prominence materials to infer the geometry of the magnetic dips.

As shown in the top row of Figure~\ref{fig6}, each arrow in the velocity fields denotes $V_{POS}$ of localized prominence materials. The value distribution ($|v_x| \ge$ 2 km s$^{-1}$, $|v_y| \ge$ 2 km s$^{-1}$) of $V_{POS}$ are roughly in alignment with the blue shift profile of oscillatory materials in Dopplergrams, as delineated by the black contours in the bottom row of Figure~\ref{fig6}.
This indicates that $V_{POS}$ and $V_{LOS}$ were almost in phase, suggesting they were projections of LALOs in POS and LOS, rather than simultaneous transverse and longitudinal oscillations.
The average $V_{POS}$ and $V_{LOS}$ of oscillating materials obtained from velocity fields and Dopplergrams, as well as the derived $V_{3D}$, were plotted in Figure~\ref{fig4}(b1) with blue, red, and black curves, respectively,
while the time-velocity curves in Figure~\ref{fig4}(a1,b1) both present a decreasing tendency.
Similarly, the angle $\alpha$ can also be obtained from Equation~\ref{eqn-3} by using average $V_{POS}$ and $V_{LOS}$,
and the values are determined to range from 5$^\circ \mbox{--}$ 15$^\circ$, which are plotted in Figure~\ref{fig4}(b2) with black curves.
Thus, the angle $\alpha$ obtained from the average $V_{POS}$ and $V_{LOS}$ also indicated that the prominence oscillation was longitudinal.

Furthermore, the angle $\beta$ in Figure~\ref{fig5} can be estimated by the average value of $arctan( |v_x|/|v_y|)$ of all arrows.
Accordingly, the values of $\beta$ are determined to decrease from $30^\circ$ to $5^\circ$ during 12:25 - 12:45 UT.
Thus, using the angle $\beta$ and average velocities ($V_{POS}$ and $V_{LOS}$ in Figure~\ref{fig4}(b1)),
the curvature radius can be calculated from Equation~\ref{eqn-8} by applying the angular velocity calculation for non-uniform time intervals (Appendix 2.3),
and the values obtained are plotted in Figure~\ref{fig4}(b2),
which can be seen that the curvature radius decreases form $\sim$190 Mm to $\sim$20 Mm, then increases to $\sim$40 Mm.

The two red curves in Figure~\ref{fig4}(a2,b2) show differences in their trends during 12:25 - 12:45 UT.
By combining Figure~\ref{fig2} and Figure~\ref{fig6},
it can be found that most of the oscillating signals in the velocity fields and Dopplergrams originated from OS1 during 12:25 - 12:45 UT,
while the start time of OS1 and OS2 was 12:15 UT and 12:25 UT, respectively.
Therefore, the red curve in Figure~\ref{fig4}(b2) displays the evolution of the partial curvature radius between the middle and top of the sinusoidal magnetic dip supporting OS1, of which the trend is similar to that in Figure~\ref{fig4}(b1).

To clearly demonstrate the 3D geometries of magnetic dips supporting OS2 and OS1,
the motion trajectories obtained from $V_{3D}$ during the first quarter periods are plotted in Figure~\ref{fig7}.
3D geometry of magnetic dip supporting OS2 in Figure~\ref{fig7}(a) exhibits a distinct sinusoidal structure,
while the blue curve in Figure~\ref{fig7}(b) exhibits a reduction in the inclination relative to the positive direction of the POS-Y axis,
which is indicative of a concomitant decrease in curvature radius from the middle to the top of the sinusoidal magnetic dip supporting OS1.

%%%%%%%%%%%%%%%%%%%%%%%%%%%%%%%%%%%%%%%%%%%%%%%%%%%%%%%%%%summary
\section{Summary and Discussion}\label{sec:sum}

In this paper, for the first time, we report the end-view observations of large-amplitude longitudinal oscillations of a quiescent prominence at the northwest limb recorded by SDO and CHASE on 2023 December 04.
The prominence oscillation was triggered by an adjoining hot structure associated with two coronal jets.
Particularly, both Doppler velocities and POS motions were detected during the oscillation.
In detail, the POS motions involved two groups of oscillation signals with similar oscillatory parameters,
an initial amplitude of $\sim$23.5 (19.5) Mm and a velocity amplitude of $\sim$30 (24) km s$^{-1}$,
each lasting for $\sim$4 cycles with a period of $\sim$77 minutes, but a phase difference of $\sim\pi/8$.
Combining the Doppler blueshift velocities, the 3D oscillatory initial amplitude and velocity amplitude are determined to be $\sim$40 Mm and $\sim$48 km s$^{-1}$,
while the angle between 3D velocities and the prominence axis ranged from 10$^\circ\mbox{--} 30^\circ$.
Apart from the time-distance diagrams,
the velocity fields obtained from optical flow method combined with Dopplergrams are also applied to calculate the curvature radius of magnetic dips,
and similar value ranges and tendency are obtained, increasing from $\sim$90 Mm at the bottom to $\sim$220 Mm in the middle,
then decreasing to $\sim$10 Mm, with transverse magnetic field strength $\ge$25 G.
Therefore, the realistic 3D geometry of the prominence magnetic dips are determined to be sinusoidal.

The triggering mechanism of prominence oscillations is an important issue.
Generally, LALOs of prominences are often triggered by energetic events near their footpoints,
such as flares or sub-flares \citep[e.g.,][]{jing2003}, coronal jets \citep[e.g.,][]{luna2014},
and sometimes caused by the impact of shock waves \citep{shen2014b} or the merging of two filaments \citep{luna2017}.
However, in this event, the oscillation was triggered by the collision and heating of an adjoining hot structure, which has been rarely reported.
As the remote brightening of hot structure occurred within cold-dense prominence material, only a little brightening materials deviated from its initial position,
while most cold-dense material did not immediately undergo significant displacement.
But once the hot plasma inside the hot structure propagated and collided with the cold-dense prominence materials, LALOs fully started.
Meanwhile, the heating process, accompanied by remote brightening, which lasted for several minutes, also caused the increase of gas pressure inside the flux rope.
Similar LALOs triggering mechanism of a jet has been described by \cite{luna2021},
they found that the collision of jet plasma and the increase of gas pressure drive prominence materials deviation with roughly equal contributions.
However, the model could not reproduce the brightening observed in cold prominence materials,
which must result from interactions involving hot-structure plasma rather than jet plasma.
Similarly, \cite{luna2024} also reported LALOs triggered by pressure imbalance arising from the interaction between the hot flare plasma and prominence materials,
involving brightening of the cold plasma.
Therefore, we could conclude that the triggering mechanism of LALOs is combined with the collision of hot plasma propagating inside the hot structure
and the increase of gas pressure inside the flux rope at the side of the brightening.

As is known, LALOs and LATOs are two oscillation modes and correspond to different seismology applied to infer the magnetic field of prominences \citep{Arregui2018}.
Thus, confirming the oscillatory direction is an essential step.
However, since LALOs and LATOS are both involved in lateral displacement \citep{Tripathi2009},
it could be controversial when investigating whether LAOs are longitudinal or simultaneous longitudinal and transverse from top view due to the low-resolution data \citep{pant2016,chen2017}.
Based on unambiguous observations \citep{zhang2017a,dai2021,tan2023}, the simultaneous LALOs and LATOs of a prominence normally exhibit different periods and phases.
In this event, the oscillation observed from end view was also involved simultaneous POS and LOS motions,
while the region of Doppler blue shift and POS motions are almost overlapping with the same phase and period of 77 minutes.
Therefore, we believe the oscillation here is purely longitudinal.
Precisely, the angles between the 3D oscillatory direction and the prominence axis in this event are determined to be $10^\circ-30^\circ$,
which is consistent with the values of LALOs detected in previous reports \citep[e.g.,][]{luna2014,zhang2017b}.

The simplified pendulum model \citep{luna2012,lunak2012,zhang2012AA} based on observed parameters of LALOs has been widely applied to infer the curvature radius and magnetic field strength of dips supporting oscillating materials.
Accordingly, the curvature radius and magnetic field strength obtained in this study were 146 Mm and 22 Gauss.
Apart from this, we also apply the non-uniform gravity pendulum model \citep{luna2022} to diagnose the related parameters as 185 Mm and 25 Gauss.
Obviously, there exists a substantial difference between the two models.
Non-uniform gravity has been theoretically demonstrated to be a significant modification to the pendulum model and still valid in applications to the non-circular magnetic dips.
Therefore, the physical parameters inferred from the non-uniform gravity model are more realistic and reasonable.

In addition to the prominence seismology, other dynamical processes in filaments, such as the skewness of mass drainage \citep{chen2014}, helical motions of threads\citep{okamoto2016}, and the counter-streaming motions \citep[e.g.,][]{wang2018}, also provide evidence to infer the configuration of their host magnetic fields.
For example, \citep{Awasthi2019} revealed the configuration of a double-decker filament by rotation motions around the filament spine and co-existing LALOs.
In this study, assuming that the magnetic field did not deform during the oscillation,
the geometry of the supporting magnetic dips,
which could be reflected by the corresponding trajectory of the motions,
was determined to be sinusoidal by integrating the oscillating 3D velocity
The results obtained from OS1 and OS2 using two methods are almost mutually verified,
which demonstrates that the method employed in this paper for calculating the curvature radius
using the differential of velocities is practical and realistic in observational studies.
Specifically, the curvature radius at the bottom of the magnetic dip was calculated to be $\sim$90 Mm.
According to the nonuniform gravity pendulum model \citep{luna2022},
the pendulum periods are concluded to be determined by the curvature radius at the bottom of the dip.
Thus, the curvature radius of 90 Mm corresponds to a corrected pendulum period of 60 minutes, while we measured an observational period of 77 minutes.
For minor differences in the two periods, the effects of the gas pressure gradient could be excluded,
as the associated slow-mode contribution would result in a pendulum period slightly less than 60 minutes.
We consider the relative computational error arising from non-uniform time intervals to be the dominant factor.
Additionally, nonlinear effects are significant since the scale of the oscillating materials is non-negligible relative to the curvature radius,
and the center of mass of the oscillating materials also changes with its deformation.

In conclusion, we propose an exercisable methodology to diagnose the 3D geometry of magnetic dips supporting oscillating materials, which is an effective supplement and improvement to prominence seismology. The realistic geometry is suggested to be sinusoidal.
This finding would need to be verified by further observations or statistical studies, and would also serve as an important reference for the geometry of magnetic dips in simulation studies.

%%%%%%%%%%%%%%%%%%%%%%%%%%%%%%%%%%%%%%%%%%%%%%%%ackonwledgements
\section{acknowledgements}

We thank the CHASE and SDO teams for providing data.
The observation data is from CHASE mission supported by China National Space Administration.
SDO is a mission of NASA\rq{}s Living With a Star Program.
AIA data are courtesy of the NASA/SDO science teams.

The authors are supported by the Special Research Assistant Project of Chinese Academy of Sciences,
the Project funded by China Postdoctoral Science Foundation (2023M733734, 2025M773193),
the Jiangsu Funding Program for Excellent Postdoctoral Talent, the Natural Science Foundation of Jiangsu Province (Grant No.BK20241706),
the Yunnan Fundamental Research Projects (Grant No. 202301AT070349), and the National Science Foundation of China (NSFC) under grants 12203097, 12403064.

Finally, I am profoundly grateful to my wife, Shengwei, and my son, Yiming, for their love and unwavering supporting throughout my postdoctoral studies in Japan.
I also thank my friend, Yafu, for his companionship during my overseas study experience.

%%%%%%%%%%%%%%%%%%%%%%%%%%%%%%%%%%%%%%%%%%%%%%%%%%%%appendix
\appendix

\section*{1. Derivation of 3D Curvature Radius Formula}

\subsection*{1.1 Basic Definitions}

Let a three-dimensional curve be parameterized by arc length $s$:
\[
\mathbf{r}(s) = (x(s), y(s), z(s))
\]
The unit tangent vector is defined as:
\[
\mathbf{T}(s) = \frac{d\mathbf{r}}{ds}
\]
The curvature $\kappa$ is defined as the magnitude of the derivative of the tangent vector with respect to arc length:
\[
\kappa = \left\| \frac{d\mathbf{T}}{ds} \right\|
\]
The radius of curvature $\rho$ is the reciprocal of the curvature:
\[
\rho = \frac{1}{\kappa}
\]

\subsection*{1.2. Representing the Tangent Vector Using Orientation Angles}

We introduce angular parameters in Figure~\ref{fig5}:
\begin{itemize}
    \item Heading angle $\psi$ (azimuth): The angle between the projection of the tangent vector onto the horizontal plane ($xy$-plane) and a reference direction (e.g., y-axes).
    \item Pitch angle $\theta$ (elevation): The angle between the tangent vector and the horizontal plane.
\end{itemize}

In a right-handed coordinate system (e.g., $y$-axis upward in Figure~\ref{fig5}), the components of the unit tangent vector $\mathbf{T}$ are:
\[
\mathbf{T} = \begin{pmatrix}
\cos\theta \cos\psi \\
\cos\theta \sin\psi \\
\sin\theta
\end{pmatrix}
\]

\subsection*{1.3. Calculating the Derivative of the Tangent Vector}

Differentiating $\mathbf{T}$ with respect to $s$, where $\psi$ and $\theta$ are functions of arc length $s$:
\[
\frac{d\mathbf{T}}{ds} = \frac{\partial \mathbf{T}}{\partial \psi} \frac{d\psi}{ds} + \frac{\partial \mathbf{T}}{\partial \theta} \frac{d\theta}{ds}
\]

Computing the partial derivatives:
\[
\frac{\partial \mathbf{T}}{\partial \psi} =
\begin{pmatrix}
-\cos\theta \sin\psi \\
\cos\theta \cos\psi \\
0
\end{pmatrix}, \quad
\frac{\partial \mathbf{T}}{\partial \theta} =
\begin{pmatrix}
-\sin\theta \cos\psi \\
-\sin\theta \sin\psi \\
\cos\theta
\end{pmatrix}
\]

Letting $\dot{\psi} = \frac{d\psi}{ds}$ and $\dot{\theta} = \frac{d\theta}{ds}$, we substitute:
\[
\frac{d\mathbf{T}}{ds} =
\dot{\psi} \begin{pmatrix}
-\cos\theta \sin\psi \\
\cos\theta \cos\psi \\
0
\end{pmatrix} +
\dot{\theta} \begin{pmatrix}
-\sin\theta \cos\psi \\
-\sin\theta \sin\psi \\
\cos\theta
\end{pmatrix}
= \begin{pmatrix}
-\dot{\psi} \cos\theta \sin\psi - \dot{\theta} \sin\theta \cos\psi \\
\dot{\psi} \cos\theta \cos\psi - \dot{\theta} \sin\theta \sin\psi \\
\dot{\theta} \cos\theta
\end{pmatrix}
\]

\subsection*{1.4. Calculating the Curvature}
We compute $\left\| \frac{d\mathbf{T}}{ds} \right\|^2$:
\begin{align*}
\left\| \frac{d\mathbf{T}}{ds} \right\|^2 &=
\left( -\dot{\psi} \cos\theta \sin\psi - \dot{\theta} \sin\theta \cos\psi \right)^2 \\
&\quad + \left( \dot{\psi} \cos\theta \cos\psi - \dot{\theta} \sin\theta \sin\psi \right)^2 \\
&\quad + \left( \dot{\theta} \cos\theta \right)^2
\end{align*}

Expanding and simplifying the first two terms:
\begin{align*}
&\left( -\dot{\psi} \cos\theta \sin\psi - \dot{\theta} \sin\theta \cos\psi \right)^2 + \left( \dot{\psi} \cos\theta \cos\psi - \dot{\theta} \sin\theta \sin\psi \right)^2 \\
&= \dot{\psi}^2 \cos^2\theta (\sin^2\psi + \cos^2\psi) + \dot{\theta}^2 \sin^2\theta (\cos^2\psi + \sin^2\psi) \\
&= \dot{\psi}^2 \cos^2\theta + \dot{\theta}^2 \sin^2\theta
\end{align*}

Adding the third term:
\[
\left( \dot{\theta} \cos\theta \right)^2 = \dot{\theta}^2 \cos^2\theta
\]

Therefore:
\[
\left\| \frac{d\mathbf{T}}{ds} \right\|^2 = \dot{\psi}^2 \cos^2\theta + \dot{\theta}^2 \sin^2\theta + \dot{\theta}^2 \cos^2\theta = \dot{\psi}^2 \cos^2\theta + \dot{\theta}^2
\]

The curvature is:
\[
\kappa = \left\| \frac{d\mathbf{T}}{ds} \right\| = \sqrt{ \left( \frac{d\psi}{ds} \right)^2 \cos^2\theta + \left( \frac{d\theta}{ds} \right)^2 }
\]

\subsection*{1.5. Formula for Radius of Curvature}

The radius of curvature is:
\[
\rho = \frac{1}{\kappa} = \frac{1}{ \sqrt{ \left( \frac{d\psi}{ds} \right)^2 \cos^2\theta + \left( \frac{d\theta}{ds} \right)^2 } }
\]

This is the formula for the three-dimensional radius of curvature expressed in terms of the heading angle $\psi$ and pitch angle $\theta$, where $\frac{d\psi}{ds}$ is the rate of change of heading angle and $\frac{d\theta}{ds}$ is the rate of change of pitch angle.

\section*{2. Derivation and application of 2D Curvature Radius Formula}

\subsection*{2.1 Definition of 2D Curvature Radius}

In planar motion, the curvature radius $R$ is related to the angular velocity $\omega$ and linear speed $v$ by:
\[
R = \frac{v}{\omega}
\]
where:
\begin{itemize}
\item $v$ is the instantaneous speed magnitude
\item $\omega$ is the angular velocity (rate of change of motion direction), $|\omega|$ approaches zero (straight-line motion).
\end{itemize}

Angular velocity $\omega$ is the derivative of the heading angle $\psi$ with respect to time:
\[
\omega = \frac{d\psi}{dt}
\]

where $\psi$ is the heading angle, defined as the angle between the velocity vector and a reference direction (typically the positive x-axis).

\subsection*{2.2 Derivation from 3D to 2D Curvature Radius Formula}\label{app22}

For a curve constrained to a two-dimensional plane:
\begin{itemize}
    \item The azimuth angle $\psi$ in Figure~\ref{fig5} is constant.
    \item The derivative of azimuth angle with respect to arc length vanishes: $\frac{d\psi}{ds} = 0$
\end{itemize}

Substitute $\frac{d\psi}{ds} = 0$ into the 3D formula and simplify:
\[
R = \frac{1}{\left| \frac{d\theta}{ds} \right|}
\]
where the physical meaning of the angle $\theta$ has also changed,
here it denotes the heading angle within a constant two-dimensional plane of $V_{POS}$ and $V_{LOS}$.

This derivation verifies that the 2D curvature formula is a special case of the 3D formula under planar constraints, demonstrating the consistency and universality of the mathematical theory.

\subsection*{2.3 application of 2D Curvature Radius Formula}

Given velocity components $V_{POS}$ and $V_{LOS}$ in Figure~\ref{fig5}, $V_{3D} = \sqrt{V_{POS}^2 + V_{LOS}^2}$ is the speed magnitude.
while the angle $\theta$ is:
\[
\theta = \arctan\left(\frac{V_{LOS}}{V_{POS}}\right)
\]

\subsubsection*{2.3.1 Angular Velocity Calculation for Non-uniform Time Intervals}
For non-uniform time interval data, we use a weighted average to calculate angular velocity (The error is approximately ):
\[
\omega_i = w_p \cdot \frac{\Delta\theta_p}{\Delta t_p} + w_n \cdot \frac{\Delta\theta_n}{\Delta t_n}
\]
where:
\begin{align*}
\Delta\theta_p &= \theta_i - \theta_{i-1} \quad \text{(normalized)} \\
\Delta\theta_n &= \theta_{i+1} - \theta_i \quad \text{(normalized)} \\
\Delta t_p &= t_i - t_{i-1} \\
\Delta t_n &= t_{i+1} - t_i \\
w_p &= \frac{\Delta t_n}{\Delta t_p + \Delta t_n} \\
w_n &= \frac{\Delta t_p}{\Delta t_p + \Delta t_n}
\end{align*}

\subsubsection*{2.3.2 Special Treatment of Endpoints}
\begin{itemize}
\item First point: Use forward difference $\omega_0 = \dfrac{\theta_1 - \theta_0}{t_1 - t_0}$
\item Last point: Use backward difference $\omega_n = \dfrac{\theta_n - \theta_{n-1}}{t_n - t_{n-1}}$
\end{itemize}
Due to the oscillation in this paper exhibiting high velocity values and low angular velocity values, smoothing preprocessing is required to reduce the differential error.

The above derivation process mainly refers to the classic textbooks of differential geometry and the online resource 'https://mathworld.wolfram.com/Curvature.html'.

\end{document}